\documentstyle[preprint,aps]{revtex}

\begin{document}

\draft

\title{Oscillatory instability of crack propagations
in  quasi-static  fracture}

\author{Shin-ichi Sasa}
\address{Department of physics, Kyoto University, Kyoto 606, Japan}

\author{Ken Sekimoto}
\address{Department of applied physics, Nagoya University,
Nagoya 464-01, Japan}

\date{\today}
\maketitle

\begin{abstract}

Crack propagations in  quasi-static fracture are  studied theoretically.
The Griffith theory is applied to discuss a crack extension condition
and motion of   crack tips in   straight propagations.
Stability of the straight propagations is investigated
based on the simple assumption that a curvature near the  crack tip
is determined by a singular shear stress.
It is shown that   straight propagations  become  unstable
under ceatain  conditions. Combined with boundary effects,
this instability leads to an oscillatory propagation of cracks.
The  critical conditions  are  calculated quantitatively.

\end{abstract}



Fracture has been   studied extensively
since  Griffith wrote a breakthrough  paper in 1920 \cite{Griffith}, but
still  provides us with many unsolved problems
\cite{Herrmann}\cite{Swinny}.
Yuse and Sano have recently carried out a nice  experiment
reminiscent of directional solidification \cite{Yuse}.
In their experiment,  a thin glass plate seeded with a small crack is
moved  from a hot region to a cold one at a velocity $v$.
It was observed that  various forms of crack patterns developed
 depending on the
temperature difference $\delta T$ and the velocity $v$.
When the temperature difference is increased
and reaches  a critical value, the crack begins to extend along a line.
At a  larger temperature difference, a  transition
from  the straight crack  to a wavy crack occurs.
In this letter,  we present  a  theory  in which
the two critical temperature differences are calculated.

An experimental configuration is displayed in fig.\ \ref{fig1}.
Here, a coordinate system $(x,y)$ is fixed on the glass plate,
which is assumed to be sufficiently thin,
to have  boundaries at $y=\pm W/2$,
to be  submerged in a cold bath
in the region $-\infty< x <s$, and
to contact  a heating element  at  $x\rightarrow \infty$.
We note that $s$  is  regarded as a variable specifying the time-dependent
external condition and that the propagation velocity of  cracks
is close to  $ds/dt$.
This fact  implies  that the crack tip moves  adiabatically with respect to
the  change of the external condition.
We thus parametrize  the time-evolution of the  position of the
crack tip by $s$ as  $(q(s),p(s))$.
The problem we address is then to derive an equation of motion for
$(q(s),p(s))$.


Suppose that a crack is positioned on the center line of
a glass plate.
According to the Griffith theory \cite{Griffith},
when the strain-energy
release rate $G$ exceeds the energy needed to create a new surface $\Gamma$,
the  crack begins to extend  at a speed of on the  order of
propagation of sound.
For a fixed temperature profile, however,
the crack would be arrested immediately at a position
where $G$ equals to $\Gamma$,
because elastic forces arising  from  non-uniform thermal expansions
are localized over a region  where the temperature gradient is
large.  Then, if a temperature profile were translated infinitesimally
along the plate, the crack  tip  would move  because  $G>\Gamma$, and then
stop again at $G=\Gamma$ after infinitesimal extension.
Since the  translation velocity $v$ is much  smaller than
the sound velocity in
glass, the  crack  propagates so as to maintain the equality $G=\Gamma$.
Such a fracture is called 'equilibrium fracture' or
'quasi-static fracture'. This should be
distinguished from an 'unstable fracture' or 'dynamic fracture' in which
the energy release rate $G$ becomes larger as the crack extends.
 In quasi-static fracture, the equation $G=\Gamma$  gives
not only  a condition for  the extension of a crack, but also the position
of a tip of the straight crack.

In order to embody this idea, we  focus on infinitely narrow cracks.
Then,  stresses  $\sigma_{ij}$  diverge at the  crack tip.
Using  a polar coordinate system $(r, \theta)$,
we can expand $\sigma_{ij}$  in $\sqrt r$ around the crack tip.
In particular, the  singular part of the stress is  characterized by
stress intensity  factors $K_I$ and $K_{II}$ related with
basic 'modes' of crack-surface displacement called mode $I$ (opening mode)
and mode $II$ (sliding mode), and its form is expressed by
\begin{equation}
\sigma_{ij}^{(sing)}={K_I \over \sqrt{2 \pi r}} \Sigma_{ij}^{(1)}(\theta)
            +{K_{II} \over \sqrt{2 \pi r}} \Sigma_{ij}^{(2)}(\theta),
\end{equation}
where $\Sigma_{ij}^{(1)}(\theta)$ and $\Sigma_{ij}^{(2)}(\theta)$
are universal functions
\cite{Freund}.
Since the shear stress vanishes  at the center of glass plates,
the equality $K_{II}=0$ holds for the straight propagation along the center
line. Then, as shown by Irwin,
 the energy release rate $G$ is related to  the
stress intensity factor $K_I$  as $G=K_I^2/E$, where $E$ is
a Young modulus \cite{Irwin}.
By introducing a critical stress intensity factor $K_I^c$
such that  $K_I^c=\sqrt{\Gamma E}$, the equation $G=\Gamma$ is reduced to
\begin{equation}
K_I=K_I^c,
\label{eqn:1}
\end{equation}
where  $K_I$ is determined by the crack pattern and  temperature profile,
and $K_I^c$ is intrinsic to the  material in question.
Noting  that $K_I$ is a function of $q(s)-s$ due to
 translational symmetry in the $x$-direction and
that $K_I$ does not depend on $s$ because  of (\ref{eqn:1}), we obtain
\begin{equation}
q(s)=s+q_0,
\label{eqn:q}
\end{equation}
where $q_0$ is a constant given by  (\ref{eqn:1}).
A crack can not extend when a solution of (\ref{eqn:q}) does not exist.
The  expression of $K_I$ for a straight  crack pattern
will be given below, but
before giving this derivation, we will discuss  an instability of
the straight propagation in the next  two paragraphs.


To this point,
the system has been assumed to be under pure mode I loading.
However, once the crack is inclined or away from the center,
a shear mode is superposed onto mode $I$,
and the crack is deflected  from  straight line geometry.
We assume here that
the crack extension condition (\ref{eqn:1}) remains  unchanged
when  the  crack extends smoothly.
Then, in order to describe  the dynamics of tips of curved cracks,
a relation between shear stresses and curvatures at the crack tips
should be considered.
Our hypothesis is as follows: a curvature  at a tip
is determined by only $K_{II}$ and the crack tip tends
 toward the orientation of minimum shear loading.
Since the curvature is well approximated by $d^2p/ds^2$ for a slightly
curved crack,  dimensional analysis leads to a simple
expression:
\begin{equation}
{d^2 p \over ds^2} = {1\over r_0}\Phi( {K_{II}\over K_I^c}),
\label{eqn:scaling}
\end{equation}
where the non-dimensional function $\Phi(z)$ is  odd in $z$ and
negative for $z>0$, and a new parameter $r_0$  with a length dimension
has been introduced.  A similar equation was considered by
R.V. Gol'dstein and R.L. Salganic \cite{Goldstein}.
They identified  $r_0$ as the length of the  end region of cracks
and derived the equation $K_{II}=0$
 by taking  the limit $r_0 dp^2/ds^2 \rightarrow 0$.
Our investigation is not limited to this limit. In fact,
we will now derive the dependence of the crack oscillation on $r_0$.

Since $K_{II}(s)$ depends on the crack pattern
which is identical to  the history of the crack tip,
$(q(s^\prime), p(s^\prime))$,
$(-\infty \le s^\prime  \le s)$,
(\ref{eqn:scaling}) is regarded as a non-Markov dynamical system.
We thus need to  consider perturbations  with infinite degrees of freedom
so as to investigate stability of  the straight propagation of cracks.
However,
if instabilities are caused by perturbations with a longer curvature
radius than the size of a most relevant region to $K_{II}$ near the tip,
we are allowed to employ  a Markov approximation.
Perturbations are then restricted to  a linear combination of
'tilt' and 'translation' of the crack, and  $K_{II}$ is written as
\begin{equation}
K_{II}=K_{II}^{(1)}{dp\over ds}+K_{II}^{(0)}p,
\label{eqn:kex}
\end{equation}
where the first and second term of the right-hand side correspond
to the stress intensity factors  of shear stresses coming from
'tilt' and 'translation', respectively.
Substituting  (\ref{eqn:kex}) into (\ref{eqn:scaling}), we
obtain a linearized equation around the straight propagation:
\begin{equation}
 {d^2 p \over d s^2}=-{1\over r_0 K_I^c}
 (K_{II}^{(1)}{dp \over ds}+K_{II}^{(0)}p),
\end{equation}
where we have used a normalization of $\Phi$ such that
$\Phi^\prime(0)=-1$.
The oscillatory instability observed in  experiments
then can be explained by checking the change of the sign of
$K_{II}^{(1)}$ and the positivity of  $K_{II}^{(0)}$.
The critical point is given by $K_{II}^{(1)}=0$
and therefore  does not depend on the unknown parameter $r_0$,
while the wavelength of oscillation $\lambda$
is proportional to $\sqrt{r_0}$:
\begin{equation}
\lambda=2\pi\sqrt{ r_0 K_I^c \over K_{II}^{(0)} }.
\label{eqn:hatyo}
\end{equation}
Our theoretical result about the
critical point can be compared to that in experiments
and the length  $r_0$ can be estimated from  experimental data.

Now,  we derive  expressions for
$K_I$, $K_{II}^{(0)}$ and $K_{II}^{(1)}$.
Since glass plates are assumed to be  sufficiently thin,
we consider  two-dimensional thermo-elastic systems
under plane stress conditions.
The equation of the force balance $\sum_j\partial_j\sigma_{ij}=0$
can be satisfied automatically when  the stresses are  written in terms of
the Airy stress function $\phi$  such that
\begin{equation}
\sigma_{xx}={\partial^2 \phi \over \partial y^2}, \qquad
\sigma_{yy}={\partial^2 \phi \over \partial x^2}, \qquad
\sigma_{xy}=-{\partial^2 \phi \over \partial x \partial y}.
\end{equation}
Then, using  Hook's law and the compatibility condition of a set of strains,
we obtain  a basic equation in our system:
\begin{equation}
\Delta^2 \phi = -E \alpha \Delta T.
\label{eqn:turiai}
\end{equation}
Here, $\alpha$ is a thermal expansion rate and
the temperature profile $T(x,t)$ along the
glass plate is  derived under  the assumptions that
$T(x,t)$ equals to $T_c$ in the cold bath $x \le s$,
and that $T(x,t)$ obeys a diffusion equation $\partial_t T= D \Delta T $
under a boundary condition $T(x,t)\rightarrow T_c+\delta T$
 for $x\rightarrow \infty$,
where $D$ is a thermal diffusion constant.
A steady moving solution at a velocity $v$ is obtained in the form
\begin{equation}
T(x,t)=T_C+\theta(x-s)\delta T(1- \exp (- (x-s)/d_0)),
\label{eqn:ondo}
\end{equation}
where we  have introduced a thermal diffusive length: $d_0=D/v$.
Stress intensity factors are given from solutions of
(\ref{eqn:turiai}) under boundary conditions
$\sum_j\sigma_{ij}n_j=0$, where $n_j$ is a normal unit vector to boundaries.
We solve the problem by dividing it into two steps.
First,
we calculate the stress $\tilde \sigma_{ij}$ in  glass plates  without
cracks, where boundary conditions at $y=\pm W/2$ and $|x| \rightarrow \infty$
are assumed as
$\tilde \sigma_{yy}=\tilde \sigma_{xy}=0$ at $ y=\pm W/2$
and
$\tilde \sigma_{xx}=\tilde \sigma_{xy}\rightarrow 0$ for
$|x|\rightarrow \infty$.
The following expressions (with  $\sigma_{ij}^{0}(x)$ denoting
 $\tilde \sigma_{ij}(x,y=0)$) are then obtained:
 \begin{equation}
\sigma_{xx}^0(x)= - E \alpha \int_{-\infty}^{\infty}dk e^{ikx}\hat T(k)f_-(k),
\label{eqn:fx}
\end{equation}
and
\begin{equation}
\sigma_{yy}^0(x)= - E \alpha \int_{-\infty}^{\infty}dk e^{ikx}
\hat T(k)(1-f_+(k)),
\label{eqn:fy}
\end{equation}
where
$\hat T(k)$ is a Fourier mode of the temperature field given
by (\ref{eqn:ondo}) and
\begin{equation}
f_{\pm}(k)={kW\cosh(kW/2)\pm 2\sinh(kW/2)\over kW+\sinh(kW)}.
\end{equation}
Then, noting that
existence of a crack  alters  {\it only} boundary conditions,
we find that
the stresses on the glass plate with a crack are  given
by $ \tilde \sigma_{ij}+\sigma_{ij}^{*}$, where $\sigma_{ij}^{*}$
is a solution of the homogeneous equation of (\ref{eqn:turiai}) under
fictious external forces introduced
so as to satisfy the boundary conditions along the crack.
Since a  singularity appears  in the stresses  $\sigma_{ij}^*$,
stress intensity factors $K_I$ and $K_{II}$ for a straight crack are
expressed by  the fictious forces such that
\cite{Mus}
\begin{equation}
K_{I}=\sqrt{2\over\pi}\int_{0}^\infty
dr{\tilde \sigma_{\theta\theta}\over\sqrt{r}},
\quad
{\rm and}
\quad
K_{II}=\sqrt{2\over\pi}\int_{0}^\infty
dr{\tilde \sigma_{\theta r}\over\sqrt{r}},
\end{equation}
where $\tilde \sigma_{\theta\theta}$ and $\tilde\sigma_{r\theta}$
correspond to  the fictious forces contributing to $K_{I}$ and $K_{II}$
respectively,
and $r$ is measured along the crack from the tip.
This formula, strictly speaking, should be  applicable to cracks
extending semi-infinitely in an infinitely large space.
We  expect however that  this approximation is still effective
for sufficiently large $W/d_0$.
By carrying out a partial integral,
we obtain the expression of  $K_{I}$ for a straight crack on the line $y=0$:
\begin{equation}
K_{I}={4\over 3}\sqrt{2\over\pi}
\int_{-\infty}^q dx (\partial_x^2\sigma_{yy}^0)(q-x)^{3/2}.
\label{eqn:k1}
\end{equation}
Also, noting a transformation $\tilde \sigma_{r\theta}
=\tilde \sigma_{xy}-
(\tilde \sigma_{xx}-\tilde \sigma_{yy})dp/ds$,
we can obtain $K_{II}$
for cracks  inclined slightly and displaced from the center:
\begin{equation}
K_{II}=\sqrt{2\over\pi}[\int_{-\infty}^q dx{\tilde\sigma_{xy}\over\sqrt{q-x}}
              +{dp\over ds} \int_{-\infty}^q dx
                     {\tilde \sigma_{yy}-\tilde\sigma_{xx}\over\sqrt{q-x}}],
\end{equation}
where the stresses have $x$-dependence in the form
$\tilde \sigma_{ij}(x,y(x))$, $y(x)=p+dp/ds\cdot (p-x)$.
We further  expand the right-hand side of this equation in
$p$ and $dp/ds$. The expansion coefficients $K_{II}^{(1)}$ and
 $K_{II}^{(0)}$ are then derived as
\begin{equation}
K_{II}^{(0)}=-2\sqrt{2\over\pi}\int_{-\infty}^q dx
             {(\partial_x^2\sigma_{xx}^0)\sqrt{q-x}},
\label{eqn:k20}
\end{equation}
and
\begin{equation}
K_{II}^{(1)}=K_I-2\sqrt{2\over\pi}\int_{-\infty}^q dx
         {(\partial_x^2\sigma_{xx}^0)\cdot(q-x)^{3/2}},
\label{eqn:k21}
\end{equation}
where we have used the equation of force balance:
$\partial_y\sigma_{xy}=-\partial_x\sigma_{xx}$.
{}From  (\ref{eqn:fx}),(\ref{eqn:fy}),
(\ref{eqn:k1}), (\ref{eqn:k20}) and (\ref{eqn:k21}),
we obtain $K_{I}$, $K_{II}^{(0)}$, and $K_{II}^{(1)}$.

We  now introduce  two control  parameters defined by
$\mu=W/d_0$ and $R=E\alpha\delta T W^{1/2}/K_I^c$.
One may easily see that $\mu$ and $R$ correspond to   non-dimensional
velocity and temperature difference, respectively.
We further define $\xi$ as a ratio of a crack length out of the cold bath
to a thermal diffusive length:
$\xi=q_0/d_0$. Then, $K_I$, $K_{II}^{(0)}$ and  $K_{II}^{(1)}$
are expressed  as
$K_I=K_I^cRF_I(\mu,\xi)$,
$K_{II}^{(0)}=K_I^cRF_{II}^{(0)}(\mu, \xi)/W$,
and
$K_{II}^{(1)}=K_I^cRF_{II}^{(1)}(\mu, \xi)$,
where $F_I$, $F_{II}^{(0)}$, $F_{II}^{(1)}$ are non-dimensional functions
calculated numerically.
When we compute  these numerical  values,
we  fix the parameter value $\mu=10$
as representaion  of ideal experimental
conditions under which the thermal diffusive length
is much larger than a thickness of glass plates.

For sufficiently small $R$, there is no solution $\xi$
satisfying   $RF_I(\mu,\xi)=1$ which is equivalent to $K_I=K_I^c$.
In this case,  the crack can not extend.
For $R>R_c^{(1)}$,
$RF_I(\mu,\xi)=1$ has a solution $\xi$
as is shown in fig.\ \ref{fig2}.
Then, when the temperature difference is increased,
the  crack length increases, and $F_{II}^{(1)}$ changes its sign
at $R_c^{(2)}$.  The straight propagation  is stable
for  $R_c^{(1)}< R < R_c^{(2)}$, but
the shear stress arising from tilt makes the crack more inclined
for $R>R_c^{(2)}$.
We  also confirmed the positivity of $F_{II}^{(0)}$.   Therefore,
we can say that a Hopf bifurcation occurs at $R_c^{(2)}$ with
a critical wavelength $\lambda_c$.
Here,  $R_c^{(1)}$, $R_c^{(2)}$ and $\lambda_c$ were
computed as $R_c^{(1)}=5.6$, $R_c^{(2)}=9.9$,
and $\lambda_c=4.5\sqrt{r_0 W}$.
Similarly,  for an arbitrary $\mu$,
assuming that  $F_I(\mu,\xi)$ takes a maximum value at $\xi_m(\mu)$ and
that $F_{II}^{(1)}(\mu,\xi)$ becomes zero at $\xi_0(\mu)$,
we obtain
\begin{equation}
R_c^{(1)}=F_I(\mu,\xi_m(\mu))^{-1},
\quad {\rm } \quad
R_c^{(2)}=F_I(\mu,\xi_0(\mu))^{-1},
\label{eqn:r12}
\end{equation}
and
\begin{equation}
\lambda_c = 2\pi \sqrt{ F_I(\mu, \xi_0(\mu))
                \over F_{II}^{(0)}(\mu,\xi_0(\mu))}
          \sqrt{r_0 W}.
\label{eqn:critl}
\end{equation}
Let us discuss  asymptotic forms of these equations
for $\mu\rightarrow \infty$.
First, from (\ref{eqn:fx}) and (\ref{eqn:fy}),
we see that the stress $\sigma^{0}$ is proportional to
$\mu$  in this limit. Then, (\ref{eqn:k1}), (\ref{eqn:k20}),  (\ref{eqn:k21}),
(\ref{eqn:r12}), and (\ref{eqn:critl})
lead to  asymptotic forms for $\mu\rightarrow \infty$:
\begin{equation}
R_c^{(1)}\sim R_c^{(2)}\sim \mu^{-1},
\qquad {\rm and} \qquad
\lambda_c \sim  \mu^{0}.
\label{eqn:rlsca}
\end{equation}
By using  dimensional quantities,
the asymptotic forms are rewritten
as
$\delta T_c \sim v^{-1} $
 and $\lambda_c \sim v^{0}$
for $v\rightarrow \infty$,  and
$\delta T_c \sim W^{-3/2}$ and $\lambda_c\sim \sqrt{W}$
for $W\rightarrow \infty$.
The latter  relation implies that the oscillatory instability is
caused by effects of the boundaries. In fact, if we assume
the periodic boundary conditions  at $y=\pm W/2$,
$K_{II}^{(0)}=0$ is obtained due to the translational symmetry  in the
$y$-direction  and the crack  then  propagates at a finite
angle to the line $y=0$ when $K_{II}^{(1)}<0$.
Similar behavior was  observed in experiments using glass  cylinders
instead of glass plates.

In the experiment done by Yuse and Sano \cite{Yuse},
the thickness of the glass plates is greater than
the thermal diffusive length in the range $v \ge 5\ [{\rm mm /s}]$.
Therefore, their experimental system may be regarded as a
two-dimensional one  only in the range $v \ll 5\ [{\rm mm /s}]$.
The relation (\ref{eqn:rlsca}) seems to be consistent with
experimental results
in the range $0.5\ [{\rm mm /s}] \le  v \ll 5\ [{\rm mm /s}]$.
In order to see correspondence with experiments more quantitatively,
we use the following values of  the material constants
typical for glass plates:
 $\alpha=8 \times 10^{-6} \ [{\rm K}^{-1}]$,
$E=7 \times 10^{10}  \ [{\rm Pa}]$,
$\Gamma=8  \ [{\rm J}\cdot{\rm  m}^{-2}]$,
$D=5 \times 10^{-3} \ [{\rm cm}^2/{\rm s}]$.
Then, for  glass plates with $W=2.4 \ [{\rm cm}]$,
$\mu=10$ corresponds to $v=0.2 \ [{\rm mm/s}]$, and the critical temperatures
$\delta T_c^{(1)}$ and $\delta T_c^{(2)}$
are computed as   $\delta T_c^{(1)}=50$ [K] and $\delta T_c^{(2)}=80$ [K].
In the experiment \cite{Yuse}, $T_c^{(1)}$ was found to be
somewhat  less than our result, while
$T_c^{(2)}$ was approximately twice of ours.
We believe that this  discrepancy may be due to error
in estimating $\Gamma$. It is not yet known how to accurately measure
this value. Further investigations will be necessary.

Finally, we  briefly discuss the parameter  $r_0$  introduced above.
{}From the experimental data that $\lambda_c=3$ [mm] for $W=1$  [cm],
we obtain  $r_0=0.05$ [mm]. As we expected above,
this value is much smaller than  curvatures  of oscillating cracks.
At this time, however,  we are  not able to satisfactly estimate the value
of $r_0$ using physical arguments.
By clarifying  what occurs on this scale, we hope to reach  a deeper
understanding of fracture phenomena.

In summary, we have proposed a theory for the transitions between the crack
patterns in quasi-static fracture. Based on our theory,
the critical points for $\mu=10$  have been  calculated quantitatively
and  the scaling relations for $\mu\rightarrow \infty$ have been  derived.
Recently, M. Marder attempted to explain the oscillatory instability
\cite{Marder} relying  on the theory of  Cotterell and Rice \cite{Cotterell}.
In addition, numerical simulations of discrete models were carried out
\cite{Taguchi}.
We have not yet been able to work out the correspondence
between these works and ours.
In order to  obtain a true understanding of this phenomena,
we need to synthesize theories, numerical simulations  and
experiments.

We are grateful to  M. Sano and A. Yuse for telling us detailed
informations of their experiments.
We thank Y-h Taguchi, Y. Hayakawa  and H. Furukawa for  communications about
their numerical simulations prior to publications. We also
thank H. Nakanishi, T. Mizuguchi, H. Hayakawa, and K. Sawada
for stimulating discussions.
G.C. Paquette are acknowledged for a critical reading of the manuscript.
This work was supported in part by the Japanese Grant-in-aid for
Science Research Fund from the Ministry of Education, Science and Culture
(No. 05740258).

\vskip5mm

\def\reflist#1#2#3#4#5{#1,{ #2}{\bf #3},  #4  (#5)}

\vskip1cm



\begin{figure}
\caption{Schematic figure of an experimental configuration.
         The  glass plate is submerged in the cold bath
         represented by the dotted region. }
\label{fig1}
\end{figure}


\begin{figure}
\caption{ $\xi$  versus $R$ (solid line) and
$K_{II}^{(1)}/K_I^{c}$  versus $R$ (dotted line).
Vertical axes of solid and dotted lines
are  graduated  on  the left and right, respectively.
These graphs starts from $R_c^{(1)}$, and $K_{II}^{(1)}/K_I^{c}$
is negative for $R \ge R_c^{(2)}$. Here,
$R_c^{(1)}=5.6$, and $R_c^{(2)}=9.9$.}

\label{fig2}
\end{figure}

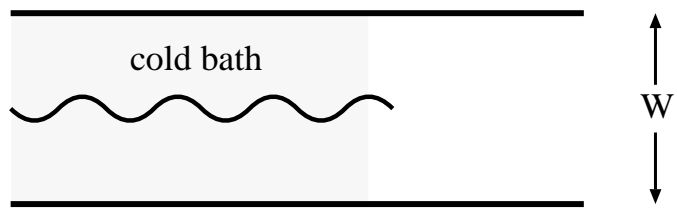
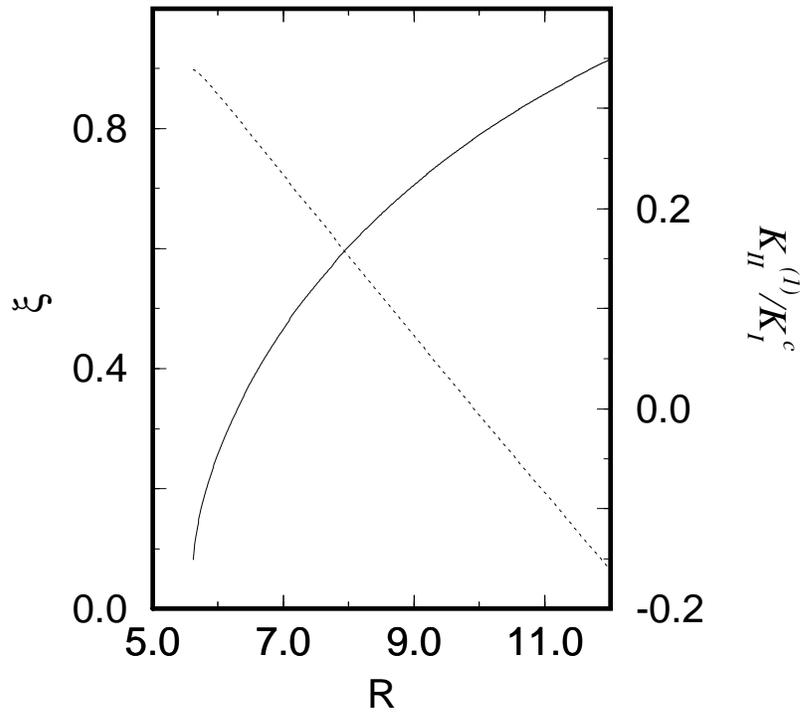
\end{document}